\documentclass[seceq]{ptptex}

\usepackage{graphicx}
\title{Higher dimensional charged BTZ-like wormhole}

\author{S. H. Hendi\thanks{email address: hendi@shirazu.ac.ir \& hendi2004@gmail.com}}

\inst{Physics Department, College of Sciences, Yasouj University,
Yasouj 75914, Iran\\
Research Institute for Astrophysics and Astronomy of Maragha
(RIAAM)-Maragha,Iran, P.O. Box 55134-441}

\abst{It is known that traversable wormhole can exist in General
Relativity only if its throat contains some exotic matter. In this
paper, we obtain $(2+1)$-dimensional (horizonless) charged
magnetic brane without (curvature) singularity. Then, we consider
a nontrivial local transformation to grant a global rotation to
spacetime. After that, we generalize magnetic brane to higher
dimensional solutions and use the cut-and-paste method to
construct higher dimensional charged BTZ-like rotating wormholes
such a way that they reduce to charged magnetic BTZ solution in
three dimensions, exactly. We also show that charged BTZ-like
wormhole supported by the exotic matter at its throat $r=r_{+}$.
Finally, we calculate the conserved quantities of the charged
BTZ-like wormhole such as mass, angular momentum and electric
charge density, and show that the electric charge depends on the
rotation parameters, interestingly, and the static wormhole does
not have a net electric charge density.}

\PTPindex{453, 450}
\begin{document}

\maketitle

\section{Introduction}

In 1935, Albert Einstein and his colleague Nathan Rosen found the
theory of intra (inter)- universe connections, so-called
Einstein-Rosen bridge \cite{EinRos1935}. After that, the name
wormhole was coined by Wheeler in his paper which discussed
wormholes in terms of topological entities called geons
\cite{Wheeler1955} and then he and Fuller proved that such a
wormhole would collapse instantly upon formation, i.e. wormhole
could not be stable \cite{Wheeler1962}.

On the other side, one of the challenging properties of wormholes
is its reversibility. We should note that traversable wormhole was
a science fiction till some of authors proposed that such a
wormhole could be made traversable by containing some form of
negative matter or energy (known as exotic matter)
\cite{Ellis,MorTho}. The type of traversable wormhole they
proposed, is referred to as a Morris-Thorne wormhole. Later, other
types of traversable wormholes were discovered as allowable
solutions to the equations of general relativity, for e.g., it was
shown that in the higher derivative gravity exotic matter is not
needed in order for wormhole to exist
\cite{Gravanis2007,HendiWormGRG,HendiWormCJP}.

In addition, three dimensional solutions of Einstein gravity are
of interest to various comprehensive issues such as gauge theory
\cite{Witten1998}, black hole thermodynamics
\cite{Carlip1995,Ashtekar2002,Sarkar2006} and string theory
\cite{Witten2007}. Also, BTZ (Banados--Teitelboim--Zanelli)
solutions in $(2+1)$-dimensions \cite{BTZ1,BTZ2,BTZ3} serves as a
worthwhile model that guides one to analyze conceptual questions
of quantum gravity as well as AdS/CFT conjecture
\cite{Witten2007,Carlip2005}.

Three dimensional wormhole solutions have been investigated in
literatures \cite{ThreeWormhole}. All of these known solutions are
uncharged. In the present paper, we find the higher dimensional
charged wormhole which its gauge potential is the same as that of
charged BTZ wormhole solutions (henceforth we call it as BTZ-like
wormhole).

Now, let us first note that the electric (Schwarzschild) gauge is
usually an appropriate choice when we are interested on black hole
solution. We should remember the fact that the electric field is
associated with the time component, $A_{t}$, of the gauge
potential while the magnetic field is associated with the angular
component $A_{\psi}$. From this facts, one can expect that a
magnetic solution can be written in a metric gauge in which the
components $g_{tt}$ and $g_{\psi \psi}$ interchange their roles
relatively to those present in the electric gauge used to describe
black hole solutions \cite{MagSol,HendiCQG}.

The outline of our paper is as follows. In next section, we
briefly present the basic field equations of the Einstein-Maxwell
gravity and discuss about properties of three dimensional static
magnetic solution. Then, we investigate a class of three
dimensional rotating solution. After that in section
\ref{HigherWorm}, we generalize our solution to arbitrary
dimensions and analyze the properties of the solutions as well as
the energy condition. In subsection \ref{Rotworm}, we endow these
spacetime with global rotations and then apply the counterterm
method to compute the conserved quantities of these solutions.
Finally, we finish our paper with some remarks.

\section{$(2+1)$-dimensional charged BTZ magnetic brane:\label{threeBTZ}}

We consider three dimensional Einstein-Maxwell theory in an asymptotically
anti de Sitter spacetime with the field equations
\begin{equation}
R_{\mu \nu }-\frac{1}{2}g_{\mu \nu }(\mathcal{R}-2\Lambda )=\left( \frac{1}{2%
}g_{\mu \nu }\mathcal{F-}2F_{\mu \lambda }F_{\nu }^{\;\lambda }\right) ,
\label{3GravEq}
\end{equation}
\begin{equation}
\partial _{\mu }\left( \sqrt{-g}F^{\mu \nu }\right) =0,  \label{3MaxEq}
\end{equation}
where $R_{\mu \nu }$ is Ricci tensor, $\mathcal{R}$ is scalar
curvature, $ \Lambda $ refers to the negative cosmological
constant which is equal to $ -1/l^{2}$. In addition, $\mathcal{F
}=F_{\mu \nu }F^{\mu \nu }$ is the Maxwell invariant, $F_{\mu \nu
}=\partial _{\mu }A_{\nu }-\partial _{\nu }A_{\mu }$ is the
Maxwell tensor and $A_{\mu } $\ is the gauge potential.

\subsection{Static Solution \label{BTZ}}

In this paper, we consider a typical metric gauge in which
$g_{tt}\varpropto -r^{2}$ and $(g_{rr})^{-1}\varpropto g_{\psi
\psi }$ instead of the Schwarzschild-like gauge in which $g_{\psi
\psi }\varpropto r^{2}$ and $(g_{rr})^{-1}\varpropto g_{tt}$
\cite{HendiBTZlike}. The most fundamental motivation comes from
the fact that we are looking for horizonless magnetic solutions
instead of electric solutions with the black hole interpretation.
The metric is given by
\begin{equation}
ds^{2}=-\frac{r^{2}}{l^{2}}dt^{2}+\frac{dr^{2}}{N(r)}+\Upsilon
^{2}l^{2}N(r)d\psi ^{2}.  \label{BTZWormMetric}
\end{equation}
It is notable that we can obtain the presented metric (\ref{BTZWormMetric})
with local transformations $t\longrightarrow il\Upsilon \psi $ and $\psi
\longrightarrow it/l$ in the known static three dimensional Schwarzschild
spacetime, $ds^{2}=-N(r)dt^{2}+dr^{2}/N(r)+r^{2}d\psi ^{2}$. Since we
changed the role of $t$ and $\psi $ coordinates, the nonzero components of
the gauge potential is $A_{\psi }$
\begin{equation}
A_{\mu }=-2ql^{2}\Upsilon ^{2}h(r)\delta _{\mu }^{\psi },  \label{AmuBTZ}
\end{equation}
where $h(r)$ is an arbitrary function of $r$. We use the
(magnetic) gauge potential ansatz, (\ref{AmuBTZ}), in the
electromagnetic field equation (\ref{3MaxEq}) and obtain
\begin{equation}
rh^{\prime \prime }(r)+h^{\prime }(r)=0,  \label{heq1}
\end{equation}%
where the prime and double primes are first and second derivative
with respect to $r$, respectively. One can show that the solution
of Eq. (\ref{heq1}) is $h(r)=\ln (\frac{r}{l})$ and the
electromagnetic field in $(2+1)$-dimensions is given by
\begin{equation}
F_{r\psi }=\frac{2ql^{2}\Upsilon ^{2}}{r}.  \label{Frpsi}
\end{equation}

To find the metric function, $N(r)$, one may use any components of
Eq. (\ref{3GravEq}). Considering the function $h(r)$, the
nontrivial independent components of the Einstein field equation,
(\ref{3GravEq}), can be simplified as
\begin{eqnarray}
\frac{N^{\prime }(r)}{r}-\frac{2}{l^{2}}+8\left( \frac{ql\Upsilon }{r}%
\right) ^{2} &=&0,  \label{E1} \\
N^{\prime \prime }(r)-\frac{2}{l^{2}}-8\left( \frac{ql\Upsilon }{r}\right)
^{2} &=&0.  \label{E2}
\end{eqnarray}%
It is notable that both $rr$ and $\psi \psi $ components of the
Einstein field equations are the same and they lead to Eq.
(\ref{E1}). After some algebraic manipulation, we find that the
solution of Eqs. (\ref{E1}) and (\ref{E2}) can be written as
\begin{equation}
N(r)=\frac{r^{2}}{l^{2}}-\left[ M+8q^{2}l^{2}\Upsilon ^{2}\ln \left( \frac{r%
}{l}\right) \right]   \label{BTZN}
\end{equation}
where $N(r)$\ is metric function and the parameters $M$ and $q$
are related to the mass and the charge of the magnetic BTZ
solution, respectively.

\subsubsection{ Geometry of the charged BTZ magnetic brane}

To investigate the geometric nature of the charged magnetic BTZ solution
given by the metric (\ref{BTZWormMetric}), we first look for curvature
singularities with their horizons. Calculation of Kretschmann and Ricci
scalars lead to
\begin{eqnarray}
R_{\mu \nu \rho \sigma }R^{\mu \nu \rho \sigma } &=&\frac{
192l^{4}q^{4}\Upsilon ^{4}}{r^{4}}-\frac{32q^{2}\Upsilon
^{2}}{r^{2}}+\frac{12}{l^{4}},  \label{3RR} \\
R &=&\frac{8l^{2}q^{2}\Upsilon ^{2}}{r^{2}}-\frac{6}{l^{2}},  \label{3R}
\end{eqnarray}
\begin{figure}[tbp]
\centerline{\includegraphics[width=10 cm,height=8 cm]
                                   {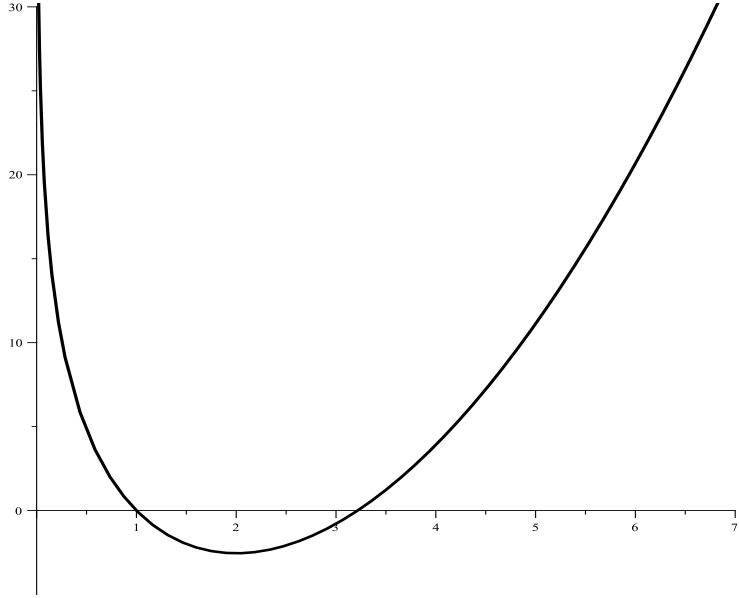}}
\caption{The metric function, Eq. (\protect\ref{BTZN}), versus $r$
for $\Upsilon =1$, $l=1$, $q=1$ and $M=1$.} \label{N(r)}
\end{figure}
which show that they diverge at $r=0$ and are finite for positive
$r$. Therefore, one might think that there is an essential
singularity located at $r=0$. Since, we are not interested in
naked singularity, we look for the existence of horizons and in
the other words, we are looking for the possible presence of
magnetically charged black hole solutions. As one can see, we will
conclude that there are no horizons and thus no black holes. The
horizons are given by the zeros of the $g^{rr}$ and so, we
investigate the case which $N(r)$ has (at least) one real positive
root ($N(r)$ has two real positive roots provided the free
parameters of the solution are chosen suitably). Taking into
account this case, the function $N(r)$ is negative for $r<r_{+}$,
and positive for $r>r_{+}$, where $r_{+}$ is the largest (real)
root of $N(r)=0$ (while $N^{\prime }(r_{+})\neq 0$, see Fig.
\ref{N(r)}). It is worthwhile to note that, $g_{rr}$ and $g_{\psi
\psi }$\ are related by $N(r)=g_{rr}^{-1}=\Upsilon
^{-2}l^{-2}g_{\psi \psi }$, and therefore when $g_{rr}$ becomes
negative (which occurs for $0<r<r_{+}$) so $g_{\psi \psi }$
becomes negative too. This fact leads to an apparent change of
signature of the metric from $(-,+,+)$ to $(-,-,-)$, and therefore
indicates that $r$ should be greater than $r_{+}$. Thus the
coordinate $r$ assumes the value $r_{+}\leq r<$\ $\infty $. The
function $N(r)$ given in Eq. (\ref{BTZN}) is positive in the whole
spacetime and is zero at $r=r_{+}$. In addition, the Kretschmann
scalar does not diverge in the range $r_{+}\leq r<\infty $. It is
notable that since $\lim_{r\rightarrow
r_{+}}\frac{1}{r-r_{+}}\sqrt{\frac{g_{\psi \psi }}{g_{rr}}}\neq
1$, this spacetime may have a conic singularity at $r=r_{+}$, in
$(r-\psi )$- section ( the same as discussed in Refs.
\cite{HendiCQG,HendiKordDoost}), and one can remove it by fix the
factor $\Upsilon =1/[lN^{\prime }(r_{+})]$ in the metric (and
obtain a three dimensional nonsingular horizonless magnetic
brane).

\subsection{Rotating BTZ magnetic brane \label{RotBTZworm}}

Now, we would like to endow our spacetime solution
(\ref{BTZWormMetric}) with a global rotation. At the first step
one may think about the mentioned local transformations
($t\longrightarrow il\Upsilon \psi $ and $\psi \longrightarrow
it/l$) in the rotating BTZ spacetime \cite{BTZ1},
$ds^{2}=-N(r)dt^{2}+dr^{2}/N(r)+r^{2}[N^{\phi }dt+d\psi ]^{2}$ and
obtain
\begin{equation}
ds^{2}=-\frac{r^{2}}{l^{2}}dt^{2}+\frac{dr^{2}}{N(r)}+\Upsilon
^{2}l^{2}[N(r)-\left( N^{\phi }\right) ^{2}]d\psi ^{2}-2N^{\phi }\Upsilon
dtd\psi .  \label{rotBTZ}
\end{equation}
Unfortunately, considering Eq. (\ref{rotBTZ}) and the field
equation (\ref{3GravEq}), lead to complicated differential
equations which we could not solve them. In order to add angular
momentum to the spacetime, we consider static metric
(\ref{BTZWormMetric}) and perform the following rotation boost in
the $t-\psi $ plane
\begin{equation}
t\mapsto \Xi t-a\psi ,\hspace{0.5cm}\psi \mapsto \Xi \psi -\frac{a}{l^{2}}t,
\label{Tr}
\end{equation}
where $a$ is a rotation parameter and $\Xi =\sqrt{1+a^{2}/l^{2}}$.
Substituting Eq. (\ref{Tr}) into Eq. (\ref{BTZWormMetric}), we obtain
\begin{equation}
ds^{2}=-\frac{r^{2}}{l^{2}}\left( \Xi dt-ad\psi \right) ^{2}+\frac{dr^{2}}{%
N(r)}+\Upsilon ^{2}l^{2}N(r)\left( \frac{a}{l^{2}}dt-\Xi d\psi \right) ^{2},
\label{Metr2}
\end{equation}
where $N(r)$ is the same as $N(r)$ given in Eq. (\ref{BTZN}).
Since the coordinate $\psi$ is periodic, the transformation
(\ref{Tr}) is not a proper coordinate transformation on the whole
manifold and therefore, the metrics (\ref{BTZWormMetric}) and
(\ref{Metr2}) can be locally mapped into each other but not
globally, and so one concludes that they are distinct \cite{Sta}.
In addition, the nonzero components of the gauge potential are
$A_{\psi }$ and $A_{t}$
\begin{equation}
A_{\mu }=2ql\Upsilon ^{2}h(r)\left( \frac{a}{l}\delta _{\mu }^{t}-\Xi
l\delta _{\mu }^{\psi }\right) ,  \label{PotBTZRot}
\end{equation}
Inserting the gauge potential ansatz, Eq. (\ref{PotBTZRot}), in
the electromagnetic field equation (\ref{3MaxEq}), surprisingly,
leads to Eq. (\ref{heq1}) with the same logarithmic form of
$h(r)$. Therefore the non-vanishing components of electromagnetic
field tensor are now given by
\begin{equation}
F_{tr}=\frac{a}{\Xi l^{2}}F_{r\psi }=\frac{2q\Upsilon ^{2}a}{r}.
\label{Ftr2}
\end{equation}
In this case, one encounters with change of metric signature for
$r<r_{+}$ and therefore we should consider this rotating spacetime
for $r>r_{+}$. This magnetic solution is both singularity-free and
horizon-less in the mentioned interval.

In addition, considering the electromagnetic field tensor given in
Eq. (\ref{Ftr2}), one can find that after applying the rotation
boost in the $t-\psi$ plane, there appears an electric field
($F_{tr}$). Here, we present a minor physical interpretation for
the appearance of the electric field. If we consider that in the
static spacetime (observer at rest), there is a static positive
charge and a spinning negative charge of equal strength, one may
conclude that this system produces no electric field since the
total electric charge is zero and the magnetic field is produced
by the angular electric current. After applying a rotation boost
to a moving observer in the static spacetime, one can show that
moving observer can see a different charge density (charge density
is a charge over a volume and this volume suffers a Lorentz
contraction in the direction of the boost) and a net electric
field appears.

\section{Generalization to higher dimensional solution:\label{HigherWorm}}

The higher dimensional action of Einstein gravity which is coupled with a
power Maxwell invariant source is given in Refs. \cite{HendiCQG,PMI}. Here,
we set power of the Maxwell invariant, $s$, to $n/2$ in the action of ($n+1$%
)-dimensional Einstein-nonlinear electromagnetic field theory \cite%
{HendiCQG,PMI}. Using the action principle, one can find the field equations
are obtained as
\begin{equation}
R_{\mu \nu }-\frac{1}{2}g_{\mu \nu }(\mathcal{R}-2\Lambda )=\alpha \left(
\alpha \mathcal{F}\right) ^{n/2-1}\left( \frac{1}{2}g_{\mu \nu }\mathcal{F-}
nF_{\mu \lambda }F_{\nu }^{\;\lambda }\right) ,  \label{GravEq}
\end{equation}
\begin{equation}
\partial _{\mu }\left( \sqrt{-g}F^{\mu \nu }\left( \alpha \mathcal{F}\right)
^{n/2-1}\right) =0,  \label{MaxEq}
\end{equation}
where negative cosmological constant, $\Lambda $, is in general
equal to $-n(n-1)/2l^{2}$ for asymptotically AdS solutions and
$\alpha $ is a constant in which we should fix it.

Here, we want to present an important motivation. It is notable
that if we generalize three dimensional charged BTZ solutions to
higher dimensional Einstein-Maxwell theory, we encounter with
basic changes in the gauge potential and also metric function. In
other word, in $(n+1)$-dimensional horizon flat static
Reissner--Nordstr\"{o}m solutions, the gauge potential and the
charge term of the metric function are proportional to
$r^{-(n-2)}$ and $r^{-2(n-2)}$, respectively. Therefore, in the
3-dimensional $(n=2)$ case, they reduce to constant values. But
for static charged BTZ solutions, the mentioned quantities are
proportional to logarithmic function of $r$. Hence, one may
conclude that in contrast with the charged BTZ black hole,
higher-dimensional Reissner--Nordstr\"{o}m solutions reduce to
uncharged solutions in three dimensions. Here, we should note that
the main goals of this paper are defining the charged magnetic BTZ
solution and then generalize our solutions to arbitrary $(n+1)$
dimensions, with wormhole interpretation, in which we will be able
to recover charged BTZ solutions, for $n=2$.

\subsection{($n+1$)-dimensional BTZ-like wormhole with a rotation parameter:\label{RotatingWorm}}

Here, we want to obtain the higher dimensional solutions of Eqs.
(\ref{GravEq}), (\ref{MaxEq}) which produce longitudinal magnetic
fields. We assume that the metric has the following form
\begin{equation}
ds^{2}=-\frac{r^{2}}{l^{2}}\left( \Xi dt-ad\psi \right) ^{2}+\frac{dr^{2}}{%
N(r)}+\Upsilon ^{2}l^{2}N(r)\left( \frac{a}{l^{2}}dt-\Xi d\psi \right)
^{2}+r^{2}d\phi ^{2}+\frac{r^{2}}{l^{2}}{{\sum_{i=1}^{n-3}}}(dx^{i})^{2}.
\label{Met1a}
\end{equation}
Note that the coordinates $-\infty <x^{i}<\infty $ have the
dimension of length and the angular coordinates $\psi $ and $\phi
$ are dimensionless with the range in $[0,2\pi]$. Also, it is
notable that one can obtain the presented metric (\ref{Met1a})
with local transformations $t\rightarrow il\Upsilon \left(
at/l^{2}-\Xi \psi \right) $ and $\psi \rightarrow i\left( \Xi
t-a\psi \right) /l$ in the horizon flat Schwarzschild-like metric,
$ds^{2}=-N(r)dt^{2}+\frac{dr^{2}}{N(r)}+r^{2}d\psi ^{2}+r^{2}d\phi
^{2}+\frac{r^{2}}{l^{2}}{{\sum_{i=1}^{n-3}}} (dx^{i})^{2}$. Here
we should note that for static case $(a=0)$, third term in Eq.
(\ref{Met1a}) vanishes and (as we will see) we need an angular
coordinate such as $\phi$ for construction of wormhole throat at
$r=r_{+}$.

Inserting the mentioned gauge potential ansatz, Eq.
(\ref{PotBTZRot}) in the electromagnetic field equation
(\ref{MaxEq}) with metric (\ref{Met1a}), surprisingly, leads to
Eq. (\ref{heq1}) with logarithmic form for $h(r)$ and the same
non-vanishing components of electromagnetic field tensor which
presented in Eq. (\ref{Ftr2}).

Now, we should fix the constant $\alpha $ in order to ensure the
real solutions. It is easy to show that for a static diagonal
magnetic metric ($a=0$) in which the only nonzero component of
$A_{\mu }$ is $A_{\psi }$, one can obtain
\[
\mathcal{F}=F_{\mu \nu }F^{\mu \nu }=8\left[ q\Upsilon lh^{\prime }(r)\right]
^{2},
\]
and so the power Maxwell invariant, $\left( \alpha
\mathcal{F}\right) ^{n/2}$, may be imaginary for negative $\alpha
$, when $n/2$ is fractional (for even dimensions). Therefore, we
set $\alpha =1$, to have real solutions without loss of
generality. Setting $\alpha =1$ with $n=2$, it is easy to show
that Eqs. (\ref{GravEq}), (\ref{MaxEq}) reduce to Eqs.
(\ref{3GravEq}), (\ref{3MaxEq}), as it should be.

To find the metric function $N(r)$, one may use an arbitrary
components of Eq. (\ref{GravEq}) such as $tt$ equation
\begin{eqnarray}
&&N^{\prime \prime }(r)+2(d-3)\frac{N^{\prime
}(r)}{r}+(d-4)(d-3)\frac{N(r)}{r^{2}}-  \nonumber \\
&&\frac{(d-1)(d-2)}{l^{2}}-2^{(d-1)/2}\left( \frac{2ql\Upsilon }{r}\right)
^{d-1}=0.  \label{tteq}
\end{eqnarray}
It is easy to find that the solution of Eq. (\ref{tteq}) (which
satisfy other components of Eq. \ref{GravEq}) can be written as
\begin{equation}
N(r)=\frac{r^{2}}{l^{2}}-\frac{1}{r^{d-3}}\left[ M+2^{(d-1)/2}\left(
2ql\Upsilon \right) ^{d-1}\ln \left( \frac{r}{l}\right) \right] ,
\label{Fhigher}
\end{equation}
where the integration constant $M$ is related to mass parameter.
One should note that these solutions are different from those
discussed in \cite{HendiBTZlike}, which were electrically charged
BTZ-like black hole solutions. The electric solutions have
BTZ-like black holes, while the magnetic solutions interpret as
BTZ-like wormhole.

\subsection{Properties of the solutions:}

In higher dimensional BTZ-like solutions the Kretschmann and Ricci scalars
are
\begin{eqnarray}
R_{\mu \nu \rho \sigma }R^{\mu \nu \rho \sigma } &=&\frac{(d-1)(d-2)^{2}(d-3)%
\left[ P\ln \left( \frac{r}{l}\right) +M\right] ^{2}}{r^{2d-2}}-  \nonumber
\\
&&\frac{2(d-2)(d-3)(2d-3)P\left[ P\ln \left( \frac{r}{l}\right) +M\right] }{%
r^{2d-2}}+  \nonumber \\
&&\frac{(4d^{2}-18d+21)P^{2}}{r^{2d-2}}-\frac{4P}{l^{2}r^{d-1}}+\frac{2d(d-1)%
}{l^{4}},  \label{RR} \\
R &=&\frac{P}{r^{d-1}}-\frac{d(d-1)}{l^{2}},  \label{R} \\
P &=&2^{(d-1)/2}\left( 2ql\Upsilon \right) ^{d-1}  \nonumber
\end{eqnarray}
which confirm that the presented solutions are asymptotically adS
and there is a curvature singularity at $r=0$. Considering $r_{+}$
as largest root of $N(r)=0$ (while $N^{\prime }(r_{+})\neq 0$), we
encounter with the same discussion about changing in metric
signature for $r<r_{+}$ and so the coordinate $r$ assumes the
value $r_{+}\leq r<\infty $. Thus the function $N(r)$ given in Eq.
(\ref{Fhigher}) is positive for $r_{+}<r<\infty $. In the same
manner, we fix $\Upsilon =1/[lN^{\prime }(r_{+})]$, to avoid conic
singularity at $r=r_{+}$ in the $(r-\psi )$-section.

Now, we investigate the wormhole interpretation of the above
solution. In order to build static wormhole ($a=0$), we may use
the cut-and-paste technique. In this construction, one should take
two copies of the solution, Eqs. (\ref{Met1a} ) and
(\ref{Fhigher}), removing from each copy the forbidden region
given by
\begin{equation}
\Omega \equiv \left\{ r\mid r<r_{+}\right\} .  \label{forbidden}
\end{equation}

With the removal of the forbidden region of each spacetime, we
obtain two geodesically incomplete spacetimes with two copies of
the following boundaries
\begin{equation}
\partial \Omega \equiv \left\{ r\mid r=r_{+}\right\} .  \label{Bforbidden}
\end{equation}
One then identifies the two copies of the mentioned boundaries
thereby obtaining a single geodesically complete manifold that
contains a wormhole joining the two regions. We can interpret the
junction $\partial \Omega $, as the throat of the wormhole. In
fact, this wormhole has a throat at $r=r_{+}$.

In order to investigate flare--out condition, we embed the 2-surface of
constant $t$, $\psi $ and $x^{i}$'s with the metric $ds^{2}=\frac{dr^{2}}{%
N(r)}+r^{2}d\phi ^{2}$, into an Euclidean flat space of one higher
dimension, which has the metric
\begin{equation}
ds^{2}=dr^{2}+r^{2}d\phi ^{2}+dz^{2},  \label{dsEuc}
\end{equation}%
The surface described by the function $z=z(r)$ satisfies
\begin{equation}
\frac{dz}{dr}=\left( \left. \sqrt{\frac{N(r)}{1-N(r)}}\right\vert
_{r=r_{+}}\right) ^{-1}\longrightarrow \infty ,  \label{dzdr}
\end{equation}
which shows that the embedded surface is vertical at the throat.
Geometrical visualization is not the only use of embedding. Since
the throat is a minimum radius from the z-axis, we know that the
embedding surface flares outward. Assuming these properties we
find that
\begin{equation}
\frac{d^{2}r}{dz^{2}}=\left. \frac{N^{\prime }}{2\left[ 1-N\right] ^{2}}%
\right\vert _{r=r_{+}}=\frac{(d-1)r_{+}}{2l^{2}}-\frac{P}{2r_{+}^{d-2}}>0,%
\text{ for }r_{+}>\left( \frac{Pl^{2}}{d-1}\right) ^{1/(d-1)}  \label{ddrdzz}
\end{equation}
which shows that the throat flare out. In other word, the
presented charged BTZ wormhole has the characteristic shape of a
wormhole, as illustrated in Figs. $1$ and $2$ of Ref.
\cite{MorTho}.

In order to hold the throat of wormhole open (stable wormhole)
there has to be a negative energy density inside. In other word,
it is shown that traversable wormholes can exist only if their
throats contain exotic matter which possesses a negative pressure
and violates the null and weak energy conditions
\cite{Ellis,MorTho,VisHoch,Exotic}. Violating the energy
conditions commits no offense against nature. Although in
classical physics the energy density of ordinary forms of matter
(fields) is believed to be non-negative \cite{Hawking}, it is a
well-known fact that energy conditions are violated by certain
quantum effects, amongst which we may refer to the squeezed vacuum
states in Maxwellian and non-Maxwellian quantum fields, Casimir
effect, gravitationally squeezed vacuum zero-point fluctuations,
classical scalar fields, the conformal anomaly, gravitational
vacuum polarization \cite{Exotic}. So perhaps exotic matter is not
utterly impossible. Undoubtedly, the issue of capturing and
storing negative energy will be left for future investigations.

Here, we discuss null and weak energy conditions for the BTZ-like
in diagonal metric (static case ($a=0$)). Wormholes which could
actually be crossed, known as traversable wormholes, would only be
possible if exotic matter with negative energy density could be
used to stabilize them. For the energy momentum tensor written in
the orthonormal contravariant basis vectors as $T^{\mu \nu
}=diag(\mu ,p_{r},p_{t_{1}},p_{t_{2}},...)$, the mathematics and
physical interpretations become simplified (this new basis is the
reference frame of a set of observers who remain always at rest in
the coordinate system and also $\widehat{g}_{\mu
\nu}=\widehat{\eta}_{\mu \nu}$). In new basis vectors, the null
energy condition (NEC) holds when $p_{r}+\mu \geq 0$ and
$p_{t_{i}}+\mu \geq 0$, and the weak energy condition (WEC)
implies $\mu>0$, $p_{r}+\mu \geq 0$ and $p_{t_{i}}+\mu \geq 0$.
The physical interpretations of $\mu $, $p_{r}$ and $p_{t_{i}}$'s
are, respectively, energy density, radial pressure and tangential
pressures that the static observers measure. For diagonal metric,
we use the orthonormal contravariant (hatted) basis vectors to
simplify interpretation
\[
\mathbf{e}_{\widehat{t}}=\frac{l}{r}\frac{\partial }{\partial t},\text{ \ \ }%
\mathbf{e}_{\widehat{r}}=f^{1/2}\frac{\partial }{\partial r},\text{ \ \ }%
\mathbf{e}_{\widehat{\psi }}=\frac{1}{\Upsilon lf^{1/2}}\frac{\partial }{%
\partial \psi },\text{ \ }\mathbf{e}_{\widehat{\phi }}=r^{-1}\frac{\partial
}{\partial \phi },\text{ \ }\mathbf{e}_{\widehat{x^{i}}}=\frac{l}{r}\frac{%
\partial }{\partial x^{i}}.
\]%
Calculations show that the stress-energy tensor is
\begin{eqnarray}
T_{_{\widehat{t}\widehat{t}}} &=&-T_{_{\widehat{\phi }\widehat{\phi }%
}}=-T_{_{\widehat{i}\widehat{i}}}=-2^{(n-2)/2}\left( \frac{F_{r\psi }^{2}}{%
\Upsilon ^{2}l^{2}}\right) ^{n/2},  \label{EMtensor1} \\
&&  \nonumber \\
\text{ \ \ }T_{_{\widehat{r}\widehat{r}}} &=&T_{_{\widehat{\psi }\widehat{%
\psi }}}=-2^{(n-2)/2}(n-1)\left( \frac{F_{\psi r}^{2}}{\Upsilon ^{2}l^{2}}%
\right) ^{n/2},  \label{EMtensor2}
\end{eqnarray}%
which one can conclude negative energy density,
$T_{_{\widehat{t}\widehat{t}}}<0$. This terminology arises because
an observer moving through the throat with sufficiently large
velocity will necessarily see a negative mass-energy density. In
addition, one can confirm the violation of NEC
\begin{equation}
T_{_{\widehat{t}\widehat{t}}}+T_{_{\widehat{r}\widehat{r}}}=T_{_{\widehat{t}%
\widehat{t}}}+T_{_{\widehat{\psi }\widehat{\psi }}}=-2^{(n-2)/2}n\left(
\frac{F_{\psi r}^{2}}{\Upsilon ^{2}l^{2}}\right) ^{n/2}<0.  \label{WEC2}
\end{equation}%
Indeed, as one expected, wormhole constitutive matter possesses
the peculiar property that its stress-energy tensor violates both
the NEC and WEC, i.e., we obtain wormhole solutions with exotic
matter at the throat.

\subsection{Wormhole solutions with more rotation parameters \label{Rotworm}}

Now, we can generalize the above solutions to the case of rotating
solutions with more than one rotation parameters. We know that the
rotation group in $(n+1)-$ dimensions is $SO(n)$ and also the
number of (independent) rotation parameters is $[n/2]$ ($[x]$ is
the integer part of $x$). The generalized rotating solution with
$k\leq \lbrack n/2]$ rotation parameters can be written as
\begin{eqnarray}
ds^{2} &=&-\frac{r^{2}}{l^{2}}\left( \Xi dt-{{\sum_{i=1}^{k}}}a_{i}d\psi
^{i}\right) ^{2}+\Upsilon ^{2}N(r)\left( \sqrt{\Xi ^{2}-1}dt-\frac{\Xi }{
\sqrt{\Xi ^{2}-1}}{{\sum_{i=1}^{k}}}a_{i}d\psi ^{i}\right) ^{2}  \nonumber \\
&&+\frac{dr^{2}}{N(r)}+\frac{r^{2}}{l^{2}(\Xi ^{2}-1)}{\sum_{i<j}^{k}}
(a_{i}d\psi _{j}-a_{j}d\psi _{i})^{2}+r^{2}d\phi ^{2}+\frac{r^{2}}{l^{2}}
dX^{2},  \label{Metr5}
\end{eqnarray}
where $\Xi =\sqrt{1+\sum_{i}^{k}a_{i}^{2}/l^{2}}$, $dX^{2}$ is the
Euclidean metric on the $(n-k-2)$-dimensional subspace and $N(r)$
is the same as $N(r)$ given in Eq. (\ref{Fhigher}). It is easy to
show that the (nonzero) components of electromagnetic field tensor
are
\begin{equation}
F_{tr}=\frac{(\Xi ^{2}-1)}{\Xi a_{i}}F_{r\psi ^{i}}=\frac{2q\Upsilon ^{2}l}{r%
}\sqrt{\Xi ^{2}-1}.
\end{equation}
Using the same approach, it is worthwhile to note that using these
solutions for $r<r_{+}$ lead to an apparent signature change and
therefore we should study this spacetime for $r_{+}<r<\infty$. So
there are neither horizon nor (curvature) singularity.

\subsubsection{Conserved Quantities \label{Conserve}}

Here, we discuss about of the values of the angular momentum, mass
density and electrical charge of the solutions. Using the
counterterm method \cite{Mal}, we consider the finite energy
momentum tensor as
\begin{equation}
T^{ab}=\frac{1}{8\pi }\left[ \left( K^{ab}-K\gamma ^{ab}\right) -\left(
\frac{n-1}{l}\right) \gamma ^{ab}\right] ,  \label{Stress}
\end{equation}
where $K^{ab}$ is the extrinsic curvature (of the boundary), $K$
is its trace, $\gamma ^{ab}$ is the induced metric (of the
boundary). To compute the conserved charges of the spacetime, we
can follow the procedure which presented in
\cite{HendiCQG,HendiKordDoost}. The first Killing vector is $\xi
=\partial /\partial t$ which is related to the total mass of the
wormhole per unit volume $V_{n-k-1} $, given by
\begin{equation}
M=\frac{(2\pi )^{k}}{4}\left[ n(\Xi ^{2}-1)+1\right] \Upsilon m.  \label{Mas}
\end{equation}
For the rotating case, the other killing vectors are $\zeta
_{i}=\partial /\partial \phi ^{i}$ which are related to the
components of angular momentum per unit volume $V_{n-k-1}$
calculated as
\begin{equation}
J_{i}=\frac{(2\pi )^{k}}{4}n\Upsilon \Xi ma_{i},  \label{Ang}
\end{equation}
which confirms that $a_{i}$'s are rotation parameters. At final
step, we should calculate the electric charge per unit volume
$V_{n-k-1}$ of the solutions. In order to calculate it one can
take into account the projections of the electromagnetic field
tensor and obtan the flux of the electromagnetic field at
infinity, yielding
\begin{equation}
Q=\frac{(2\pi )^{k}2^{3n/2}n\Upsilon \left( \frac{q}{l}\right) ^{n-1}}{16}
\sqrt{\Xi ^{2}-1}.  \label{elecch}
\end{equation}
It is worth noticing that the electric charge of the wormhole per
unit volume $V_{n-k-1}$ is proportional to the rotation
parameters, and is zero for the static wormhole solutions ($\Xi
=1$). This result is expected since now, besides the magnetic
field along the $\psi$ coordinate, there is also a radial electric
field ($F_{tr} \neq 0$) and the former component leads to an
electric charge.

\section{ Closing Remarks}

In this paper, we have demonstrated a construction
$n+1$-dimensional magnetically charged solutions, namely charged
BTZ-like wormholes. Considering three dimensional magnetic
solution of Einstein-Maxwell gravity, we obtained a static charged
BTZ magnetic brane. We found that the Kretschmann scalar is finite
in whole spacetime and there is no horizon. Then, we used an
improper coordinate transformation to add an angular momentum in
the magnetic brane.

After that, we generalized our solutions to arbitrary $(n+1)$
dimensions in which, in contrast with higher dimensional
Reissner--Nordstr\"{o}m solutions, we were able to recover BTZ
solutions, for $n=2$. We also checked the flare-out condition and
found that charged BTZ-like solutions satisfied the flare-out
condition (with wormhole interpretation) provided the free
parameters of the solution are chosen suitably. In addition,
calculation of energy conditions showed that charged BTZ-like
wormholes supported by an exotic matter at their throats
($r=r_{+}$).

Finally, we generalized charged BTZ-like wormhole solutions to the
case of rotating solutions with more rotation parameters and
calculated some conserved quantities such as angular momentum and
mass density. Also, we analyzed the flux of the electromagnetic
field at infinity for the BTZ-like solutions and found that the
electric charge density of the charged BTZ-like wormholes depends
on the magnitude of the rotation parameters, while the static
charged BTZ-like wormholes have no net electric charge.

We should note that it is worthwhile to investigate the dynamic
stability of the charged BTZ-like wormhole solutions with respect
to radial perturbations.

\section*{Acknowledgements}
This work has been supported financially by Research Institute for
Astronomy \& Astrophysics of Maragha (RIAAM).

\end{document}